\documentclass[%
 reprint,
nofootinbib,
 amsmath,amssymb,
 aps,
prb,
]{revtex4-2}

\usepackage[caption=false]{subfig}

\usepackage{graphicx}% Include figure files
\usepackage{dcolumn}% Align table columns on decimal point
\usepackage{bm}% bold math
\usepackage{slashed}
\usepackage{mathtools}
\usepackage{cleveref}
\DeclarePairedDelimiter\abs{\lvert}{\rvert}%

\begin{document}

\preprint{2411.02147}

\title{A minimalistic model for inelastic dark matter}% Force line breaks with \\%

\author{Giovani Dalla Valle Garcia}%
 \email{giovani.garcia@student.kit.edu}
\affiliation{Institut für Astroteilchenphysik, Karlsruhe Institute of Technology, Karlsruhe, Germany
}%

\date{\today}% It is always \today, today,

\begin{abstract}
 Models of inelastic (or pseudo-Dirac) dark matter commonly introduce a gauge symmetry spontaneously broken by the introduction of a dark sector version of the Higgs mechanism. We find that this ubiquitous introduction of two extra fields, a vector and a complex scalar boson, is indeed unnecessary, with only a mass generating real scalar field being actually required. We consider a simple UV-complete model realizing this minimal setup and study the decays of the excited dark matter state as well as constraints from perturbative unitarity, (in)direct detection and colliders. We find that, in the visible freeze-out scenario ($ \text{DM} \, \text{DM} \leftrightarrow \text{SM} \, \text{SM}  $), we still have unconstrained regions of parameter space for dark matter masses $\gtrsim 100$~GeV. Moreover, most of the available regions either present long-lived excited states, which are expected to interfere with the standard cosmological history, or will be probed by future direct detection experiments, such as DARWIN, due to the unavoidable residual elastic interactions. The only regions remaining out of experimental reach present highly fine-tuned parameters.
\end{abstract}

\maketitle

%\tableofcontents

\section{\label{sec:intro}Introduction}

While the nature of dark matter (DM) remains mysterious, its energy density is probed gravitationally with ever increasing precision. Particle models for DM must include some sort of production mechanism, which can explain the observed DM relic abundance. Among the possible mechanisms, thermal freeze-out won a lot of attention due to its connection to the Weakly Interacting Massive Particle (WIMP) Miracle and is particularly appealing because of its independence from the initial conditions of the universe, thanks to equilibrium physics. However, traditional WIMPs have continuously faced strong blows from direct detection experiments, and recently the focus has been shifting to searching for alternatives that keep the WIMP paradigm predictive power. 

The simple WIMP model featuring CP-conserving fermionic DM with a dark Higgs mediator~\cite{Kim:2008pp} is essentially ruled out by direct detection constraints in the visible freeze-out regime (relic abundance thermally produced via DM annihilation into Standard Model (SM) particles, $ \text{DM} \, \text{DM} \leftrightarrow \text{SM} \, \text{SM}  $)\cite{Duerr:2016tmh}. A clever solution is to replace the dark Higgs with a pseudoscalar mediator (a model also referred to as Coy Dark Matter), or alternatively, to introduce significant levels of CP-violation in the dark sector~\cite{Boehm:2014hva,Ghorbani:2014qpa,Lopez-Honorez:2012tov}. However, these approaches significantly limit the potential for investigating DM properties through direct detection experiments~\cite{Ghorbani:2018pjh,Ertas:2019dew}. Furthermore, these solutions suffer from strong indirect detection constraints for DM masses below 10--100~GeV~\cite{Baek:2017vzd}. We propose substituting the usual Dirac-Yukawa ($\bar{\chi}\chi$) interactions in the dark sector by a Majorana-Yukawa ($\bar{\chi^c}\chi$) type which are generically neglected in these singlets extensions of the Standard Model. In this case, allowing for a small parity breaking transforms the model into one of the most minimal forms of inelastic Dark Matter (iDM)~\cite{Tucker-Smith:2001myb} where indirect detection bounds can be easily avoided by depleting the heavier state abundance, e.g., via its decays.  

In this context, iDM models commonly found in the literature are usually mediated by a vector portal with the SM, essentially introducing a $U(1)$ gauge symmetry in the dark sector which is spontaneously broken via a dark Higgs mechanism, require non-renormalizable SM-portals or extend the SM with a large number of new particles~\cite{Izaguirre:2015zva,Darme:2017glc,Foguel:2024lca,Bae:2023ago,Masso:2009mu,Chang:2010en,Patra:2011aa}. We find that, by simply introducing a single real pseudoscalar field acquiring a vacuum expectation value (vev), we can construct a UV-complete fermionic iDM model with only one main assumption: a small but nonzero parity breaking in the dark sector Yukawa interactions. Given the minimalistic nature of this iDM model, we will refer to it as minimal-inelastic Dark Matter (miDM). The new model turns out to have a very predictive visible freeze-out regime, which can be nearly completely probed by future direct detection experiments, such as DARWIN, along with cosmological probes, e.g., Big Bang nucleosynthesis (BBN) predictions for the primordial element abundances or measurements of the Cosmic Microwave Background (CMB).

In this letter, we first introduce the miDM framework and elucidate its relationship with the SM. We pay particular attention to the decays of the excited DM state as well as the constraints imposed by perturbative unitarity. Subsequently, we discuss the DM relic abundance focusing on the freeze-out mechanism. Indirect and direct detection bounds are discussed from both elastic (diagonal) and inelastic (off-diagonal) interactions as well as collider constraints originating from precision measurements of the SM-Higgs properties. Finally, we present the experimental bounds combined with the relic abundance calculations along some general cosmological features and conclude.       

\section{The framework}

We consider a real pseudoscalar field $A$, with a nonzero vev $\langle A\rangle = w$, and a Dirac fermion $\chi_D = \chi_L + \chi_R$, where both fields are singlets under all SM gauge groups. We impose a $Z_4$ symmetry,\footnote{Turning the pseudoscalar complex would allow us to introduce a $U(1)$ symmetry, if gauged it would effectively reproduce the niDM model in ref.~\cite{Garcia:2024uwf}. The $U$(1) symmetry would similarly simplify our Lagrangian although an analysis of the effects of a Nambu–Goldstone mode associated with the symmetry breaking or a new gauge boson  goes beyond the scope of this work.}\begin{align}
    \chi \to i \chi && , && A \to -A \, ,
\end{align} in order to simplify the number of terms in the Lagrangian. Then, the most general Lagrangian for the new fields together with the SM fields allowed by the SM gauge symmetries and the $Z_4$ symmetry can be written as \begin{equation} \label{eq:NewPhysicsLagrangian}
    \mathcal{L}_{\text{NP}} =  \mathcal{L}_{\chi}  + \mathcal{L}_{A}  \, ,
\end{equation} where
\begin{align} 
   \mathcal{L}_{\chi}  = & \bar{\chi} (i\slashed{\partial}  - m_d)\chi  - \dfrac{1}{\sqrt{2}}y A\bar{\chi^c} (\delta_P + \gamma_5)\chi + \text{h.c.} \,, \label{eq:LFermion} \\
    \mathcal{L}_{A} = & \dfrac{1}{2}(\partial^{\mu} A)(\partial_{\mu} A)  + \dfrac{\mu_a^2}{2} {A}^2  - \dfrac{\lambda_a}{4} {A}^4 -  \dfrac{\lambda_{ah}}{2} {A}^2 \abs{H}^2 \, ,  
    \label{eq:Lscalar} 
\end{align} with the SM-Higgs field denoted by $H$ and the charge conjugated field by $\psi^c = C \gamma_0^T \psi^*$ where $C$ is the charge conjugation matrix. If $\delta_P = 0$ and $y \in \mathbb{R}$, both P and CP would be good symmetries of the Lagrangian, effectively becoming an example of Coy Dark Matter~\cite{Boehm:2014hva}. In the following, we will adopt a real and small breaking parameter $\delta_P = \delta_P^\ast \ll 1$ as well as a real Yukawa coupling $y$.\footnote{For simplicity, we take $\delta_P$ to be real, but a complex value would still have similar phenomenological implications. The only special point would be an imaginary $\delta_P^\ast=-\delta_P$ which would restore CP in $\mathcal{L}_{\chi}$ if $y \in \mathbb{R}$. } Its origins could be related to an initial fully CP-conserving dark sector where some parity breaking emerges from extra new  physics interactions of the dark sector with the SM weak interactions, where parity is maximally broken. Nevertheless, in this work, we will remain agnostic about its origin, rather we will focus on studying its consequences.

 We adopt the unitary gauge, \begin{align}
    A = a+w && \text{and} && H = \dfrac{1}{\sqrt{2}}(0,h+v)^T 
\end{align} where $v=246$~GeV is the SM-Higgs vev. Then, the fermion mass terms become \begin{align} 
    \mathcal{L}_{\chi}  \supset - m_d\bar{\chi}_L \chi_R +  \frac{1}{2} m_L \bar{\chi}^c_L \chi_L -  \frac{1}{2}m_R  \bar{\chi}^c_R \chi_R + \text{h.c.} \, ,
\end{align} where the Majorana masses are related to the Yukawa couplings and the scalar vev as $m_{L/R} = 2 y_{L/R} w$ with $y_{L/R} = y (1\mp \delta_P)/\sqrt{2} $. It is technically natural to have $w\ll m_d$, implying \begin{equation}\label{eq:deltaMdef}
    \delta_d \equiv  \frac{m_L}{m_d} \propto \frac{w}{m_d}  \ll 1 \, .
\end{equation}  

Now, in order to achieve a nearly off-diagonal coupling of the DM states with the dark pseudoscalar, following Appendix B of ref.~\cite{Garcia:2024uwf}, we simply need to adopt $ \delta_i \ll 1$. Therefore, \cref{eq:deltaMdef} together with the assumption of a small parity breaking $\delta_P \ll 1$ allows us to expand our results in both $\delta_P$ and $\delta_d$, thus, henceforth, all equations will be presented in leading order in these variables, or in next-to-leading order when we consider it pertinent. However, we maintain the full dependence on $\delta_i$ for all the numerical results.

After diagonalization of the fermionic mass terms, one finds \begin{equation}
    \mathcal{L}_{\chi} = \dfrac{1}{2} \Bar{\chi}_i  (i \slashed{\partial}  - m_i) \chi_i - \dfrac{1}{2} y_L \, a   \,\Bar{\chi}_i  ( {\alpha}_{ij} + i {\beta}_{ij}  \gamma^{5} ) \chi_j \, ,
\end{equation}  where the masses of the Majorana fermions are given by \begin{equation}
   m_{\chi^{(\ast)}} = m_d\left(1+ \frac{1}{2} \delta_d( \delta_d \mp 2 \delta_P )\right) \, ,
\end{equation} implying the normalized mass difference \begin{equation}
    \Delta_m \equiv {(m_{\chi^{\ast}}-m_{\chi})}/{m_{\chi}} = 2\delta_d \delta_P  \,.
\end{equation} The interaction coefficients are given in \cref{tab:InteractionCoefficientsScalar} where the lower-script ``$\ast$" denotes the excited state index and ``$\_$" the ground state one.
\begin{table}[h]
\centering
\caption{\label{tab:InteractionCoefficientsScalar} Coefficients for interactions between dark fermions and the dark scalar in the regime $\delta_i \ll 1$.  } 
\begin{tabular}{l l}

\hline \vspace{-0.33cm} \\ 

 $ $ $ {\alpha}_{\ast\ast} =  2 ( \delta_d + \delta_P + \Delta_m + \delta_P^2) \;\;$  &$ {\beta}_{\ast\ast} = 0$\\
  
   $ $ $ {\alpha}_{\ast\_} = 0$  &$ {\beta}_{\ast\_} =  2 + 2\delta_P +  2\delta_P^2  -\delta_d^2$  $ $\\
  
   $ $ $ {\alpha}_{\_\ast} = 0$   &$ {\beta}_{\_\ast} = {\beta}_{\ast \_}$\\
   
   $ $ $ {\alpha}_{\_\_} =  2 ( \delta_d - \delta_P + \Delta_m - \delta_P^2)$ &$ {\beta}_{\_\_} = 0$\\
\hline
\end{tabular}
\end{table}

One can define the dark Yukawa fine-structure constant and the diagonals and off-diagonal fine-structure constants as \begin{align}
    &\alpha_y = y^2/4\pi \, , \\ &\alpha_{\text{el}\_} =  \alpha_y {\alpha}_{\_\_}^2 \, ,  && \alpha_{\text{el}\ast} =  \alpha_y {\alpha}_{\ast\ast}^2 \, ,
 && \alpha_{\text{inel}} = \alpha_y {\beta}_{\ast\_}^2 \, .
\end{align} An interesting feature of the coupling structure found here is that the diagonal couplings are different for ground and excited states, just as in inelastic Dirac Dark Matter (i2DM). This makes it possible to consider an iDM model with non-negligible mediator decays into two excited states which can give rise to striking signatures such as two displaced-vertices. 

Interactions with the SM are possible due to a mass mixing of the dark pseudoscalar $A$ and the SM-Higgs $H$ generated by the $\lambda_{ah}$-term in $\mathcal{L}_A$. A rotation of the fields by an angle $\theta$, given by \begin{equation}
    \tan 2\theta \equiv \dfrac{\lambda_{ah} v w }{\lambda_h v^2 - \lambda_a w^2} \, ,
\end{equation} diagonalizes the scalar mass matrix. The rotation couples $a$ to the SM and $h$ to the dark sector with couplings suppressed by $\sin{\theta}$. The final scalar masses can be found in ref.~\cite{Duerr:2016tmh} along with more details on the full scalar Lagrangian diagonalization. For simplicity, even after the rotation, we will continue referring to the physical SM-like Higgs by $h$ and the physical dark pseudoscalar field by $a$.

\subsection{Excited Dark Matter decays}

\begin{figure*}
\subfloat[Light mass splittings.\label{fig:1a}]{%
  \includegraphics[width=0.465\textwidth]{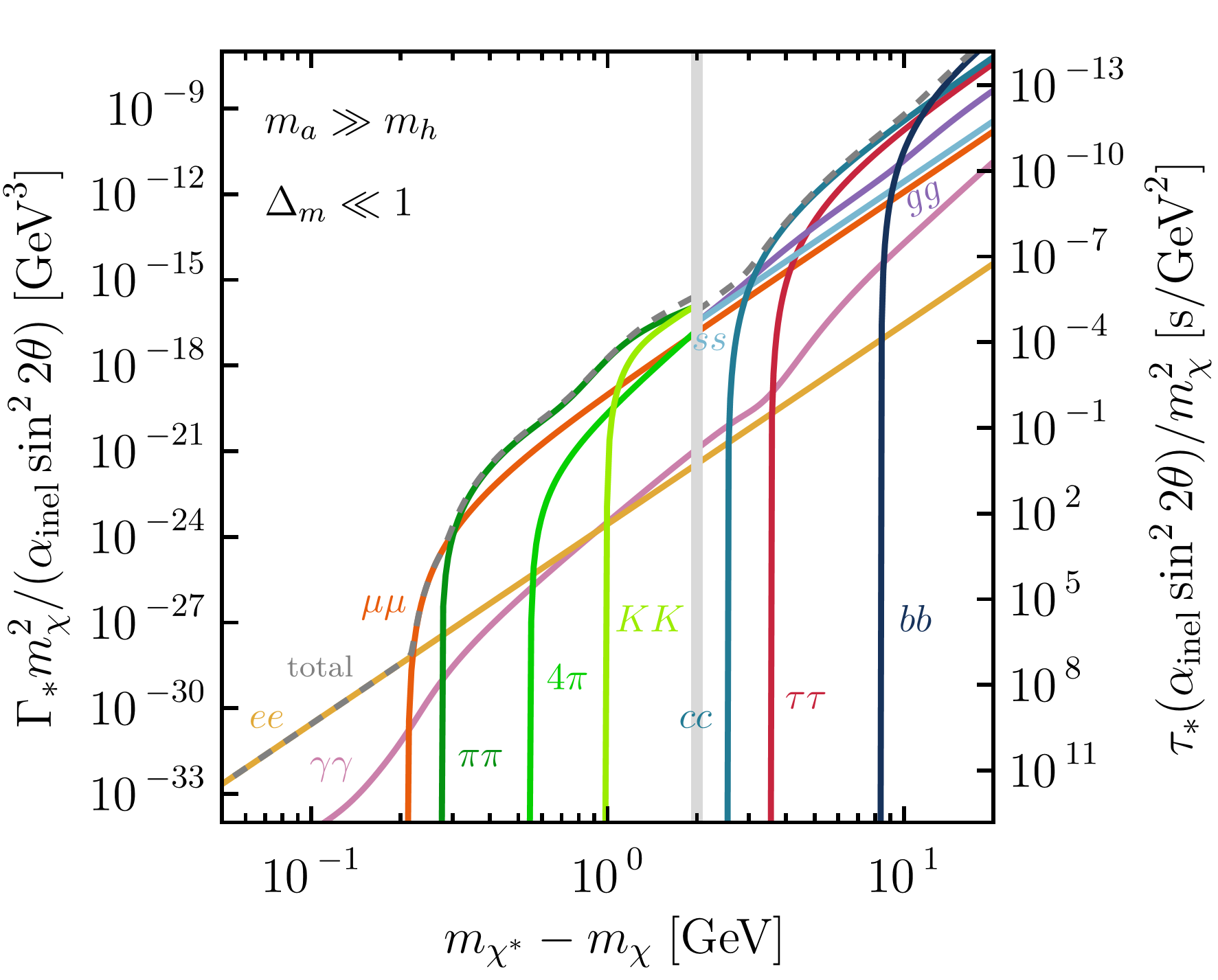}
}%
\hfill
\subfloat[Heavy mass splittings.\label{fig:1b}]{%
  \includegraphics[width=0.46\textwidth]{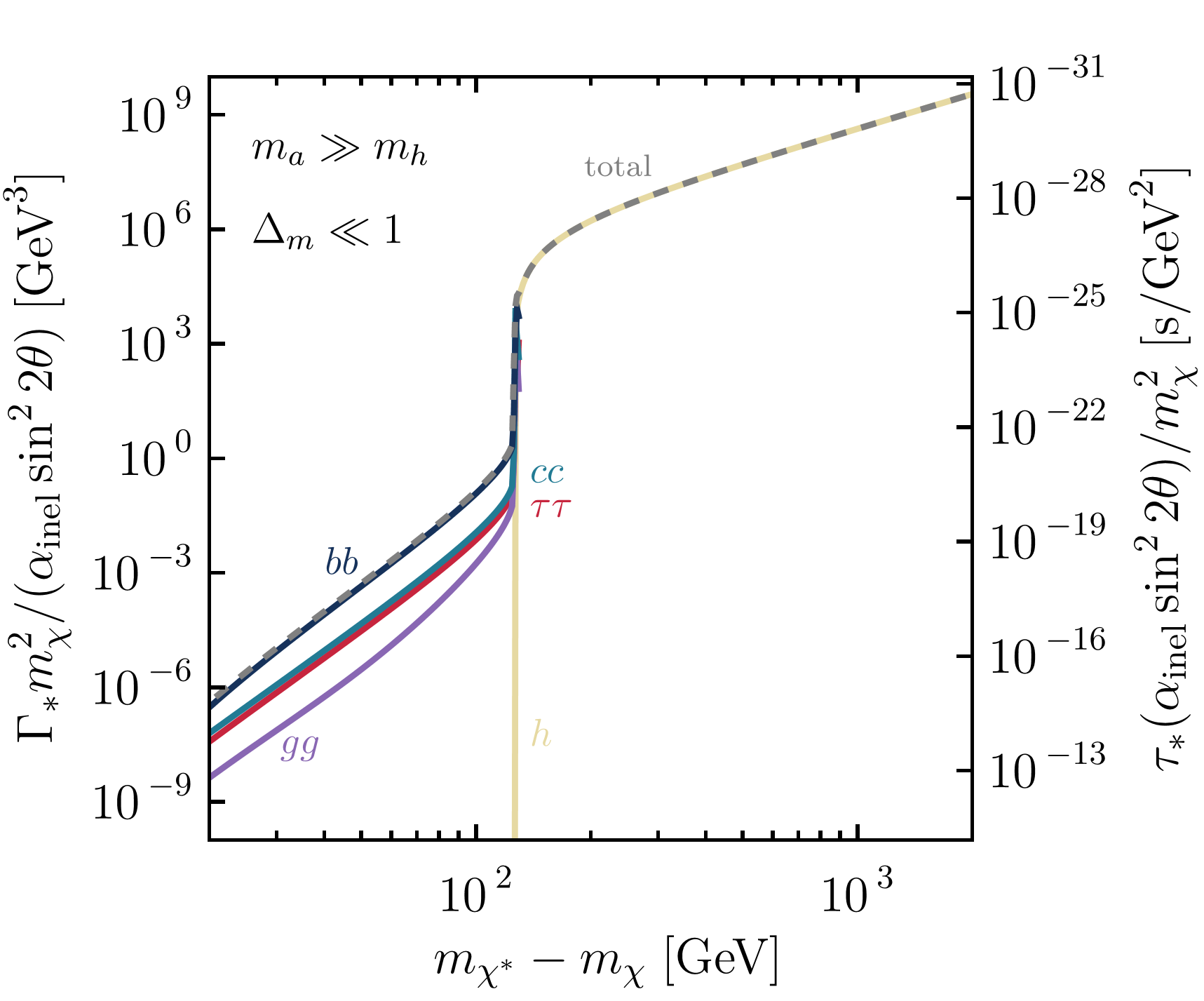}
}

\caption{\label{fig:ExcitedDecays} Total decay width of the excited DM state $\Gamma_\ast$ (dashed gray) and partial decay widths $\Gamma(\chi^\ast \to \chi \mathcal{F})$ into different SM final states $\mathcal{F}$ (solid colors) for (a) small and (b) large mass splittings. We present only dominant decay modes. The scale on the vertical axis on the right side presents the lifetime of the excited DM state $\tau_\ast$. The vertical light gray band at $\Delta_m m_\chi=$~2~GeV indicates where neither the dispersive analysis nor the perturbative spectator model are reliable~\cite{Winkler:2018qyg,Ferber:2023iso,Boiarska:2019jym}, therefore, the discontinuity in the total width. 
For decays into the Higgs boson, we approximate $\cos \theta \approx 1$.}
\end{figure*}

If kinematically allowed, decays into dark sector final states dominate the decay width of the excited state $\Gamma_\ast$ for $\theta \ll y$. However, since the mass splitting in miDM is naturally small $\Delta_m \ll 1$, $\chi^\ast \to 3\chi$ is kinematically forbidden, while $\chi^\ast \to \chi a$ is kinematically forbidden in the visible freeze-out regime ($\chi^\ast \chi \leftrightarrow \overline{\text{SM}}\text{SM}$). Thus, $\chi^\ast$ decays into the SM are, in general, of crucial importance and the contributions to $\Gamma_\ast$ of different SM decay channels are depicted in \cref{fig:ExcitedDecays}.\footnote{We note that we neglect the reduced phase space due to confinement, i.e., charm (bottom) quarks can only be created for mass splittings superior to  $2 m_D$ (2$m_B$) rather than for the simple quark masses threshold $2m_c$ ($2m_b$).}  Loop-induced decays into photons~\cite{Marciano:2011gm} and gluons~\cite{Winkler:2018qyg,Spira:1995rr} (using $\alpha_s(s)$ from \cite{ParticleDataGroup:2024cfk}) as well as non-perturbative decays into mesons~\cite{Winkler:2018qyg} are computed using results from the usual dark Higgs decays found in the literature via the amplitude formula \begin{equation}
     \overline{ \lvert \mathcal{M} \lvert^2 }(\chi^{\ast} \to \chi \,2\mathcal{F}) = \dfrac{\beta_e}{\beta_\mathcal{F}} \dfrac{\Gamma_{a^\ast\to 2\mathcal{F}}}{\Gamma_{a^\ast\to 2 e}}\overline{ \lvert \mathcal{M} \lvert^2 }(\chi^{\ast} \to \chi \,2e) \, ,
\end{equation} where $2\mathcal{F}$ stands for any SM particle-antiparticle pair and $\beta^2 _\mathcal{F}= {m_{2\mathcal{F}}^2-4m_{\mathcal{F}}^2}$ with $m_{P_1 P_2}$ being the invariant mass of the pair of particles $P_1 P_2$.\footnote{As shown in ref.~\cite{Ovchynnikov:2023von}, the hadronic decays suffer from large uncertainties which should be taken into account in case one intends to study the cosmology of the miDM model in detail. } Here, the decay width $\Gamma_{a^\ast}$ corresponds to the off-shell width of the pseudoscalar $a$ with an effective mass given by $m_{2\mathcal{F}}$. 

In the perturbative regime, the decay width into light fermions, $2m_f < \Delta_m m_\chi \ll m_\chi$,  is simply given by \begin{multline}
    \Gamma_\ast (\chi^\ast\to\chi \Bar{f}f) \approx \\ N_c \dfrac{\alpha_{\text{inel}} \sin ^22 \theta } {560 \pi^2 } \dfrac{m_f^2}{v^2} \dfrac{\left(\Delta_m^2  - \tau_{f}^2\right)^4}{\Delta_m  + \tau_{f}}  \left(\dfrac{1}{m_h^2}-\dfrac{1}{m_a^2}\right)^2  m_\chi^5\, ,
\end{multline} where $N_c = 3(1)$ for quarks (leptons) and $\tau_f = 2m_f/m_\chi$.

For very light dark scalar masses $m_a < \Delta_m m_\chi \ll m_\chi $, 2-body decays of the excited DM state are allowed $\chi^\ast \to \chi \, a$, consequently, for $\theta \ll y$, $\Gamma_{\ast}$ will be given by \begin{equation}
    \Gamma_{\ast} \approx \dfrac{1}{2}  \alpha_{\text{inel}}  \cos^2{\theta} \left(\Delta_m^2 - \dfrac{m_a^2}{m_\chi^2}\right)^{\frac{3}{2}} m_\chi \, .
\end{equation}

Considering DM masses in the range of $10$ to $10^4$~GeV, for relatively light dark scalars $m_a \ll m_h$ and using values for the couplings which are expected in the freeze-out regime ($\theta,y \gtrsim 10^{-1}$), one finds that the excited state decays in time intervals well below a second for mass splittings larger than a kaon pair $\Delta_m   > \tau_K$.  While for a heavy dark scalar $m_a \gg m_h$, this occurs for mass splittings above the $b$-quark mass $m_\chi\Delta_m   > m_b$. Therefore, in these regions, we naturally expect negligible cosmological abundances of the excited DM state. For $m_a \approx m_h$, the two scalar portals present destructive interference which can turn the excited state into a long-lived particle and a detailed study of the initial relic density of $\chi^\ast$ would be required.

\subsection{Perturbative Unitarity}

For DM to be produced via visible freeze-out, one needs to require $m_a \approx \sqrt{2\lambda_a}w > m_\chi \approx m_d $. While, in order to suppress diagonal couplings, we have to be placed in the parameter region where $\delta_d \ll 1$, i.e., $\sqrt{2}y_L w \ll m_d$. Therefore, we push $\lambda_a$ to large values which can break perturbativity. Following ref.~\cite{Duerr:2016tmh}, we will impose   \begin{align}  \label{eq:pertubativityBound}
    3(\lambda_h+\lambda_a) \pm \sqrt{9(\lambda_h-\lambda_a)^2 + \lambda_{ha}^2} < 16 \pi \, ,
\end{align}  in order to enforce perturbative unitarity in the processes: $a a \to a a$ and $h h \to h h $.\footnote{Stronger bounds may be obtained by also considering other processes~\cite{Kang:2013zba}.} In addition, we will also exclude parameter regions where the dark Yukawa coupling becomes non-perturbative, $\alpha_y \geq 1$. Not surprisingly, these requirements place strong limits on much of the parameter space for visible freeze-out, see the detailed results in section~\ref{sec:results}.

For some initial intuition on what values of $\delta_d$ one can expect in the visible freeze-out regime, we can make a simple estimate for a lower bound on $\delta_d$. First, we notice that for $\lambda_{ha} \approx 0$ (i.e., $\theta \approx 0$), the perturbativity bound in \cref{eq:pertubativityBound} simplifies to $\lambda_a < 8\pi/3$ and, by demanding $m_a > m_\chi$, we find that $\delta_d \gtrsim \sqrt{3/8 \pi} y$. Second, in the freeze-out approximation, one finds \begin{equation}
    \langle \sigma v \rangle \approx \dfrac{y_{\text{SM}}^2 y^2 \theta^2}{16\pi m_\chi^2}e^{-x_f\Delta_m} \approx 10^{-10} \text{ GeV}^{-2} \, ,
\end{equation} where $y_{\text{SM}}$ is the coupling of the SM-like Higgs to the SM final states, which we will take approximately to be order unity since we are interested in $m_\chi \gg $ GeV, and $x_f = m_\chi/T_f \approx 20$ with $T_f$ denoting the freeze-out temperature. Now, from the above freeze-out condition, we conclude that \begin{equation}
    \delta_d \gtrsim 0.25 \left( \dfrac{0.1}{\theta}\right) \left( \dfrac{m_\chi}{1\text{ TeV} }\right)e^{10\Delta_m} \,.
\end{equation} The above suggests that $\alpha_{\text{el}\_}$ cannot be arbitrarily  suppressed  at the visible freeze-out regime since $\alpha_{\_\_} \sim 2 \delta_d $ (except by fine-tuning $\delta_d \approx \delta_P$). Moreover, we expect relatively large mixing angles $\theta \gtrsim 0.1$ and small normalized mass splittings $\Delta_m \lesssim 0.1$ in order to avoid  perturbativity bounds. Overall, these results are just indicative since, in all numerical results, we keep the full perturbativity formulae and perform more precise relic abundance computations.

\section{Relic abundance}

In order to predict the DM relic abundance, we have developed a \texttt{Mathematica} package to solve Boltzmann equations for the particular case where we have up to two DM particles in kinetic equilibrium with the SM during freeze-out (see Appendix \ref{ap:Aboltzmann})~\cite{Evans:2017kti}. In the package, we also include a relic abundance estimator via a detailed freeze-out approximation that is also applicable for the two DM particles case. This approximation is used to create the figures. In addition, by computing the full solution for a limited number of points in each figure, we manage to reduce the approximation error in our plots. We limit our studies to $m_\chi \geq 10$~GeV in order to avoid non-perturbative QCD effects in the relic abundance calculation, which go beyond the scope of this introductory work on miDM.

\begin{figure}
     \includegraphics[width=0.4\textwidth]{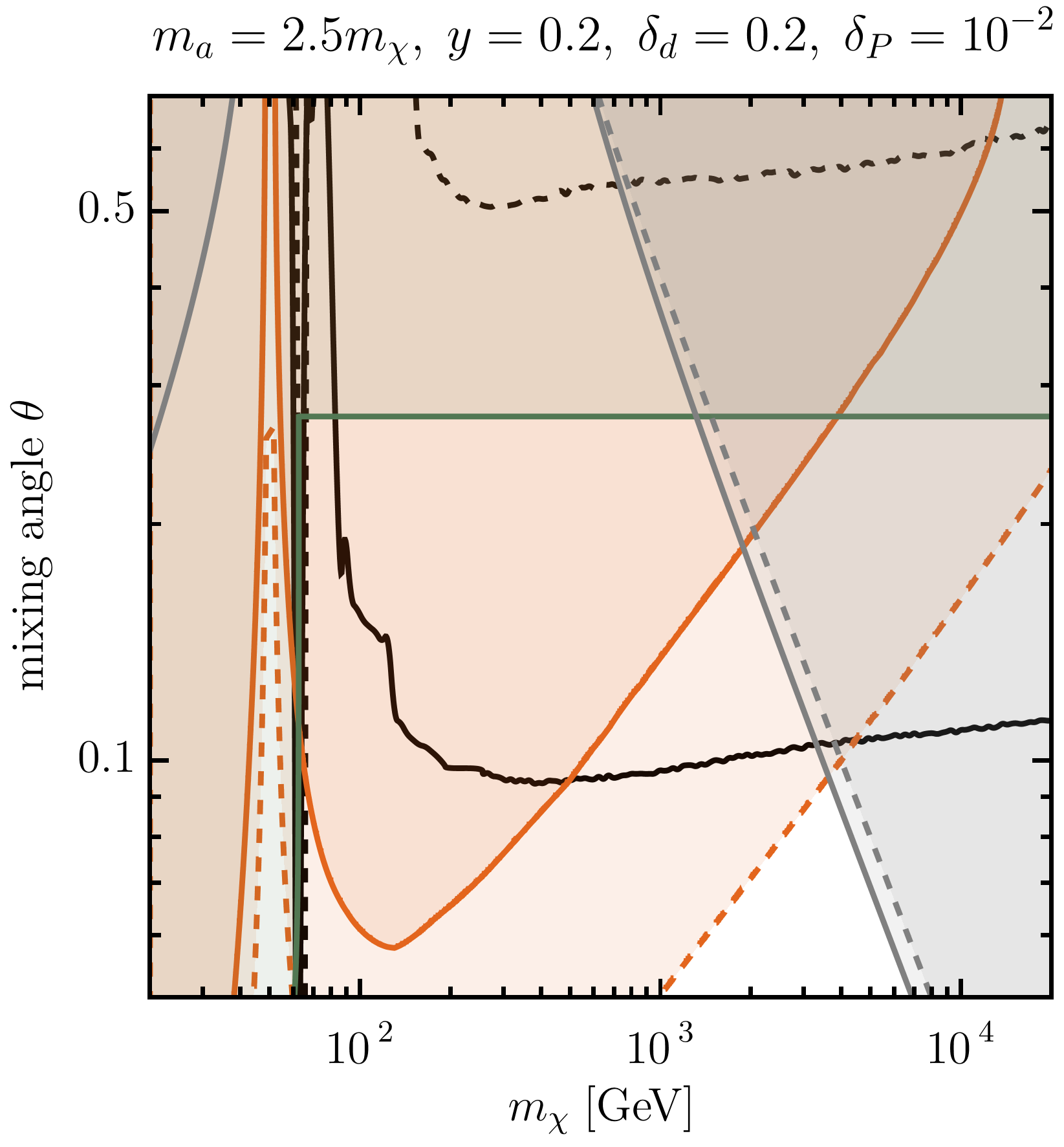}\hspace{0.25cm}
     \caption{Values of the mixing angle $\theta$ which reproduce the observed DM relic abundance $\Omega_{\rm DM}h^2 = 0.12$ as a function of the DM mass (black lines).  The gray shaded regions correspond to perturbativity bounds. Higgs signal strength constraints from CMS and ATLAS~\cite{CMS:2022dwd,ATLAS:2022vkf} are presented in green while orange regions relate to bounds from the direct detection experiment LZ~\cite{LZCollaboration:2024lux}, see the text for details. Solid lines stand for the miDM scenario while dashed ones for the classical elastic dark Higgs portal DM. }
     \label{fig:relicComparison}
 \end{figure}

 In miDM, the off-diagonal interactions lead to $s$-wave  DM co-annihilation into SM particles allowing to reproduce the correct DM relic abundance with weaker couplings compared to the usual $p$-wave scalar portal DM (a shared feature with Coy Dark Matter, i.e., pseudoscalar portal DM). For comparison, in \cref{fig:relicComparison}, we present the thermal target for both the standard $p$-wave and the miDM scenarios, where we clearly see the advantages of miDM in surviving direct detection bounds. In both models, DM is composed of Majorana fermions.

\section{(In)direct detection}

Indirect detection bounds from off-diagonal (inelastic) co-annihilation or $\chi^\ast$ decays are irrelevant in the parameter region where the lifetime of $\chi^\ast$ is much shorter than a second since the short lifetimes imply a negligible abundance of the excited state during BBN and CMB.\footnote{See e.g. ref~\cite{Gustafson:2024aom} for a potential signal from indirect detection for iDM models with large mass splitting.} For $\tau_\ast \gtrsim 0.02$~s~\cite{Boyarsky:2020dzc}, the excited state abundance is non-negligible and we leave its cosmological implications for future studies. Nonetheless, we point out that bounds from energy injections during BBN are expected to be stringent in these cases. In order to completely circumvent these bounds, one would need to consider extremely long-lived states ($\delta_P \approx 0$) where studies on late universe decaying DM would be required. On the other hand, the diagonal (elastic) annihilation is velocity-suppressed due to its $p$-wave nature, thus, leaving no relevant constraints on the annihilation cross section. Furthermore, important constraints could arise from the visible decays of the pseudoscalar mediator. However, in the mass range we are interested $m_a \geq 10$~GeV and for the mixing mixing angles we consider $\theta \gtrsim 10^{-2}$, we find $\tau_a \ll 0.02$~s implying that the mediator abundance can be neglected already during BBN.

For direct detection searches, inelastic scatterings are  suppressed either by the mass splitting being larger than the kinetic energy of the DM particles in our galaxy $\Delta_m > v^2 \sim 10^{-6}$ or by their pseudoscalar nature which only generates suppressed operators in the effective Lagrangian for direct detection, specifically $\mathcal{O}_{10,11} \propto \vec{q}$~\cite{Fitzpatrick:2012ix}.\footnote{The off-diagonal coupling also generates spin-independent DM-nucleus interactions at the loop level. However, results from ref.~\cite{Ertas:2019dew}, applicable to our model in the limit $\Delta_m \ll1$, show that direct detection constraints, considering only the off-diagonal coupling, are still nearly an order of magnitude away from reaching the thermal target line.} On the other hand, the elastic coupling induces unsurpassed spin-independent DM-nucleus interactions, which are heavily constrained by direct detection experiments. In particular, the LUX-ZEPLIN experiment~\cite{LZCollaboration:2024lux} has placed the strongest constraints to date on the cross section of this type of interaction and in the future DARWIN~\cite{DARWIN:2016hyl} will push these constraints even further. Following ref.~\cite{Duerr:2016tmh}, this cross section is given by \begin{equation}
    \sigma^{\text{SI}} = \dfrac{\mu_{\chi N}^2m_N^2 f_N^2 \alpha_{\text{el\_}} \sin^2 2\theta}{v^2}\left(\dfrac{1}{m_h^2}-\dfrac{1}{m_a^2}\right)^2 \, ,
\end{equation} where $\mu_{\chi N}$ is the reduced mass of the DM-nucleon system, $m_N$ the nucleon mass and $f_N = 0.3 $ is the effective coupling of DM to nucleons~\cite{Boveia:2016mrp}. The only way of relaxing these constraints is by requiring a $ \alpha_{\text{el\_}}$ sufficiently small. However, most of the visible freeze-out regime allowed by perturbativity is still within the reach of the above mentioned detectors.  

\section{Colliders}

We consider a Higgs signal strength modifier $\mu$ affecting equally all production and decay modes of the SM-like Higgs,  while keeping all branching fractions equal to the SM predictions. Combined measurements performed at ATLAS and CMS impose the observed $\mu$ to be higher than 0.93 at 95\% confidence level~\cite{ATLAS:2022vkf,CMS:2022dwd}. According to ref.~\cite{Bhattiprolu:2022ycf}, $\mu$ for our mass mixing case is given by \begin{equation}
    \mu = \cos^2 \theta \dfrac{\cos^2 \theta \,\Gamma_h}{\cos^2 \theta \,\Gamma_h +  \Gamma_{\text{dark}}}
\end{equation} with $\Gamma_h = 3.2$~MeV the total decay width of $h$ and $\Gamma_{\text{dark}}$ the partial decay width of $h$ into dark sector final states.  

In this work, we focus only on constraints coming from the measurement of the signal strength, as they are expected to be stronger than any other collider constraint to date for $m_a \geq 10$~GeV in the visible freeze-out regime (where $a$ essentially decays invisibly)~\cite{Ferber:2023iso}.  However, patches of the secluded regime ($ \text{DM} \, \text{DM} \leftrightarrow a \, a  $) can be probed by searches at ATLAS~\cite{ATLAS:2018sbw,ATLAS:2022vkf}, CMS~\cite{CMS:2018amk} and LEP~\cite{LEPWorkingGroupforHiggsbosonsearches:2003ing}, showing the complementarity of all current  searches for new physics and an exciting potential for future data from the HL-LHC.

\section{Results} \label{sec:results}

\begin{figure*}
\begin{minipage}[t]{0.39\linewidth}
\includegraphics[width=\linewidth]{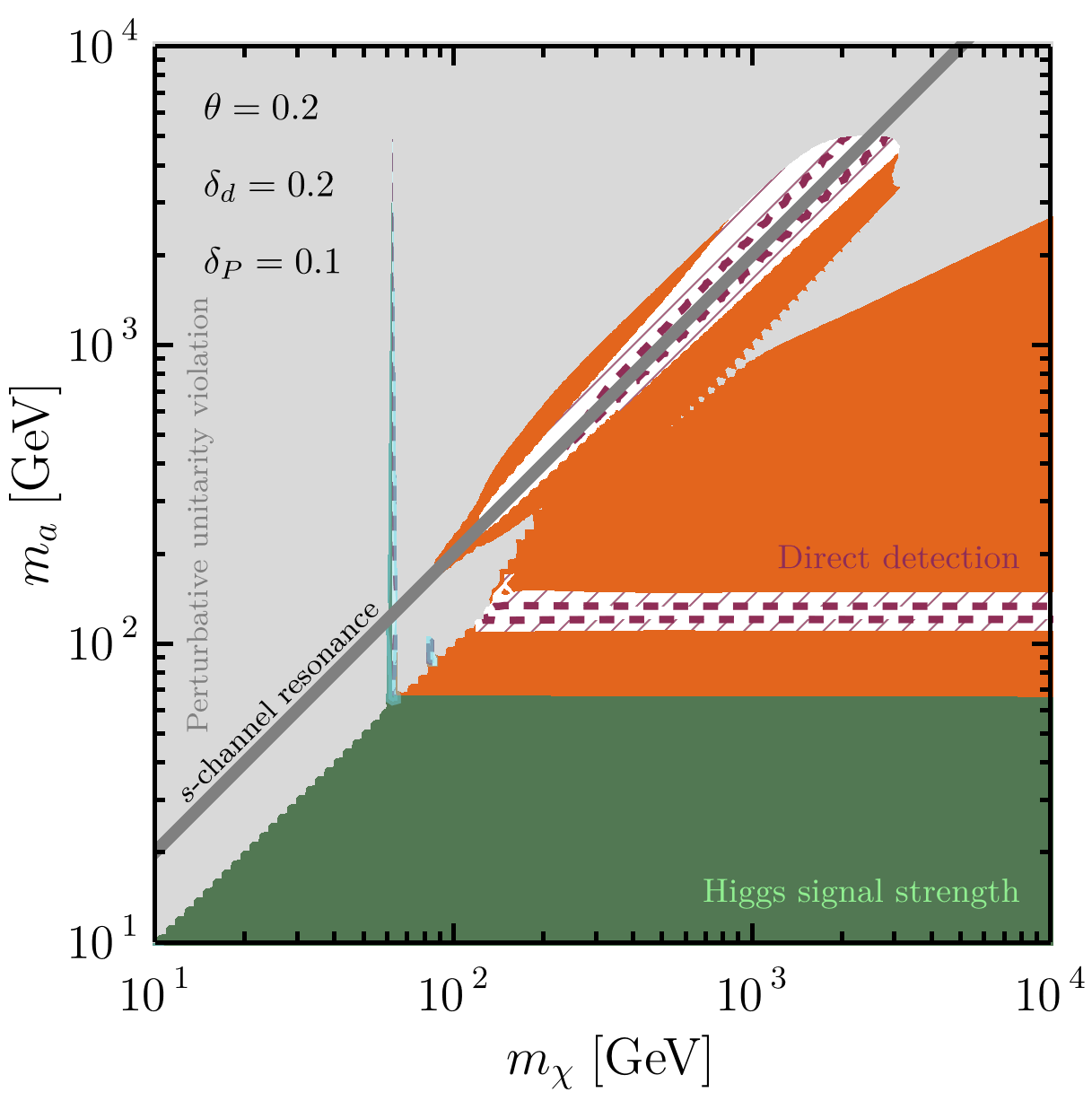}
\end{minipage}\hspace{1.25cm}%
\begin{minipage}[t]{0.39\linewidth}
\includegraphics[width=\linewidth]{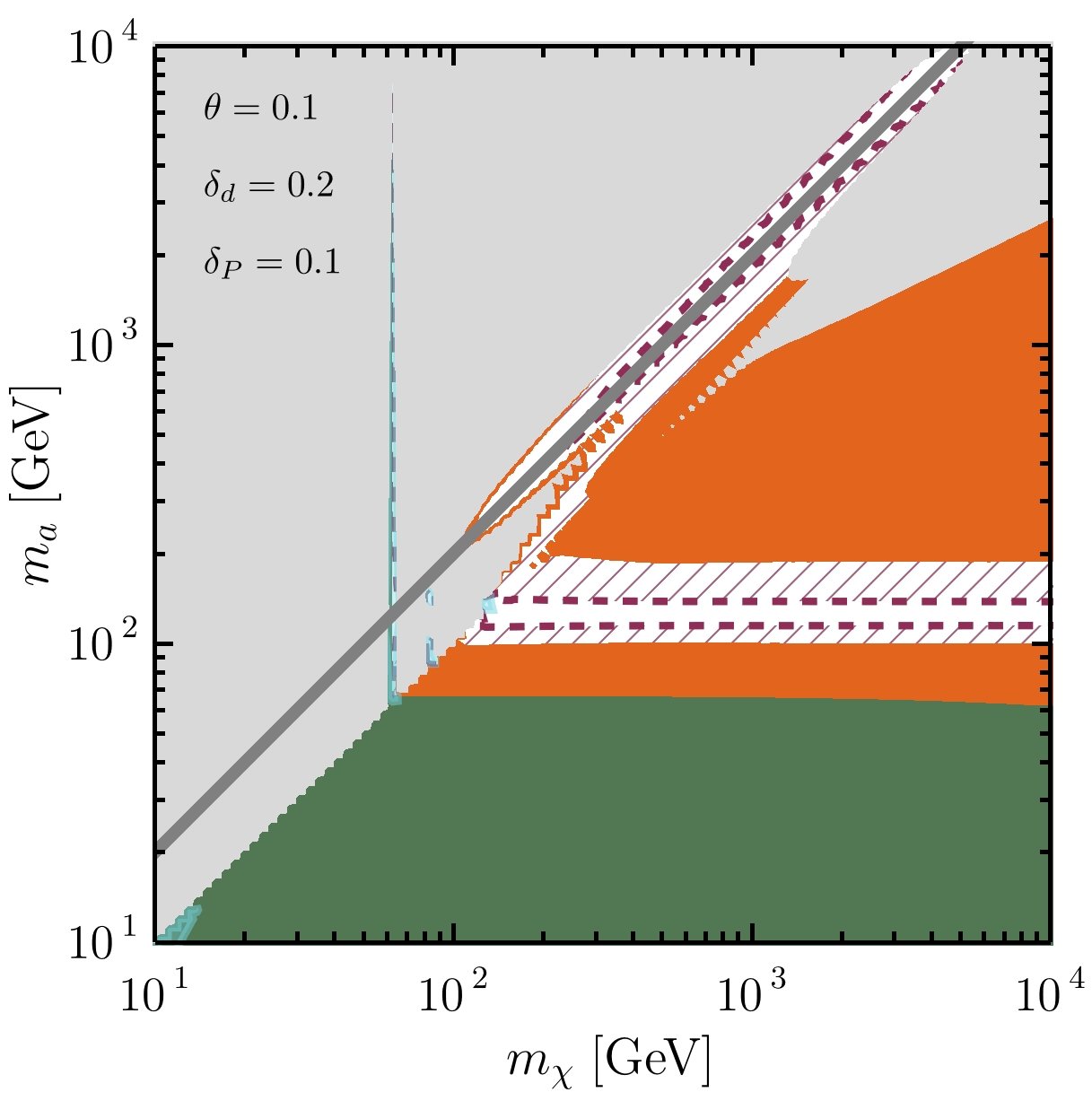}
\end{minipage}\newline%
\begin{minipage}[t]{0.39\linewidth}
\includegraphics[width=\linewidth]{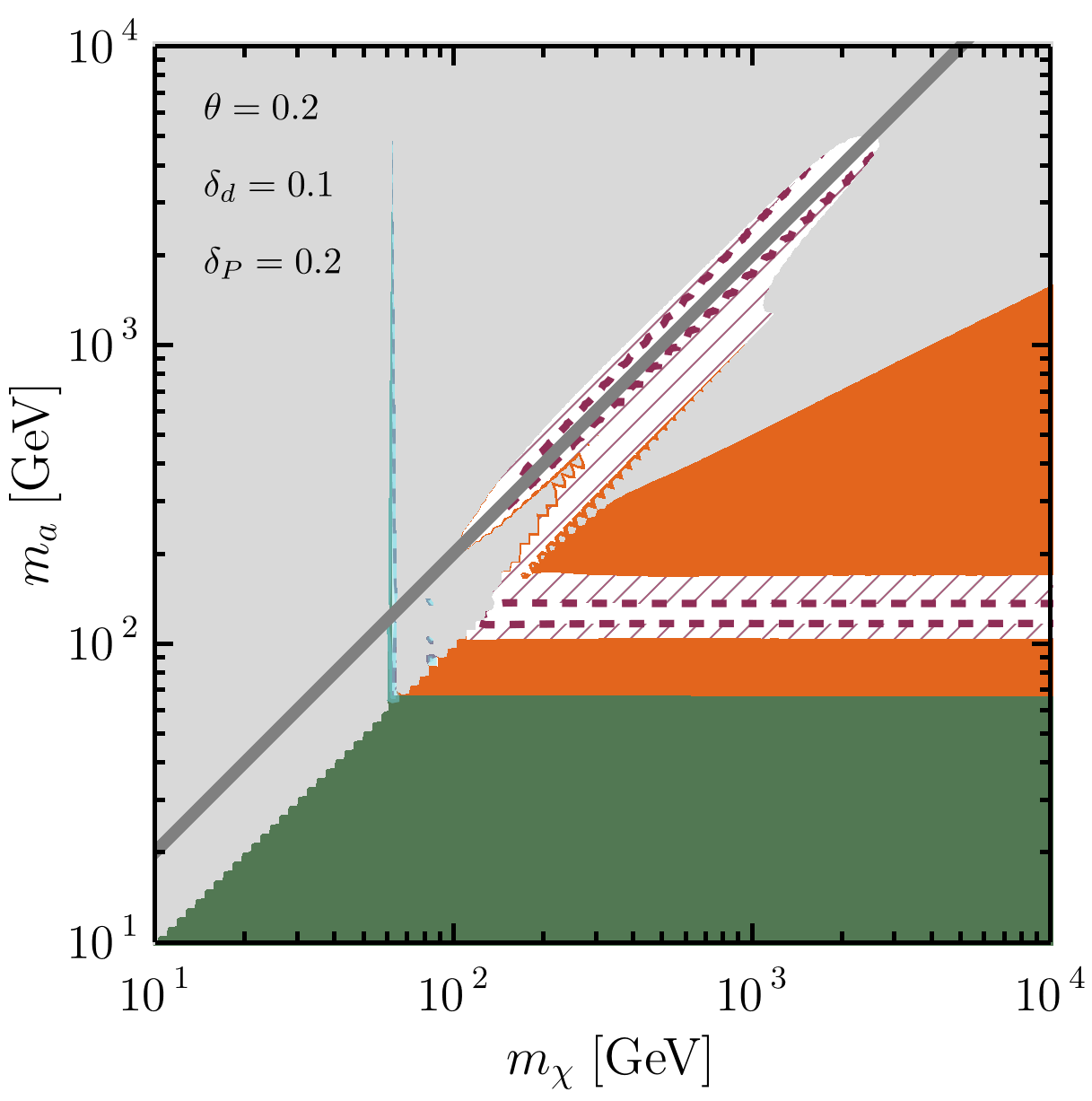}
\end{minipage}\hspace{1.25cm}%
\begin{minipage}[t]{0.39\linewidth}
\includegraphics[width=\linewidth]{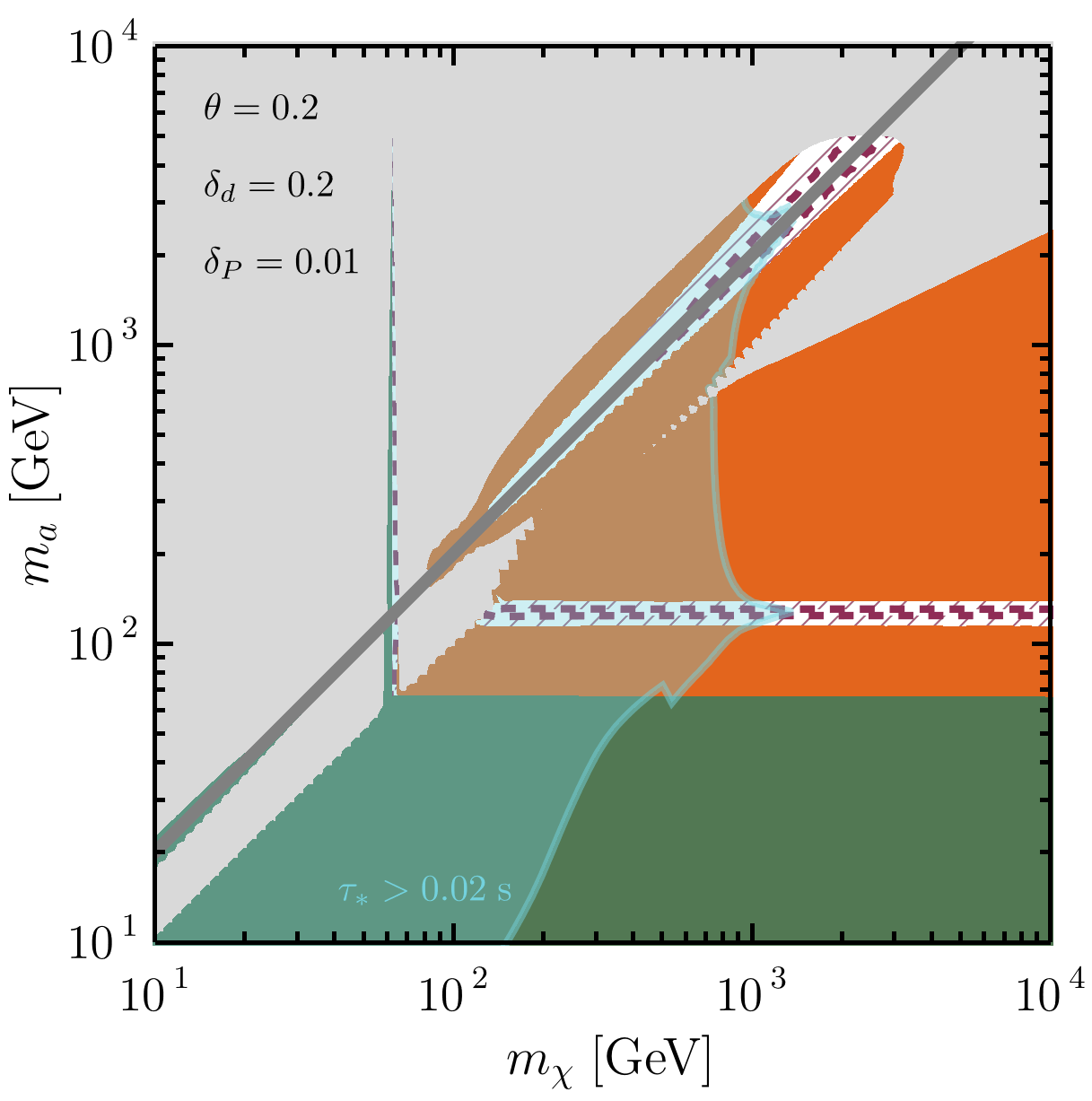}
\end{minipage}\newline%
\begin{minipage}[t]{0.39\linewidth}
\includegraphics[width=\linewidth]{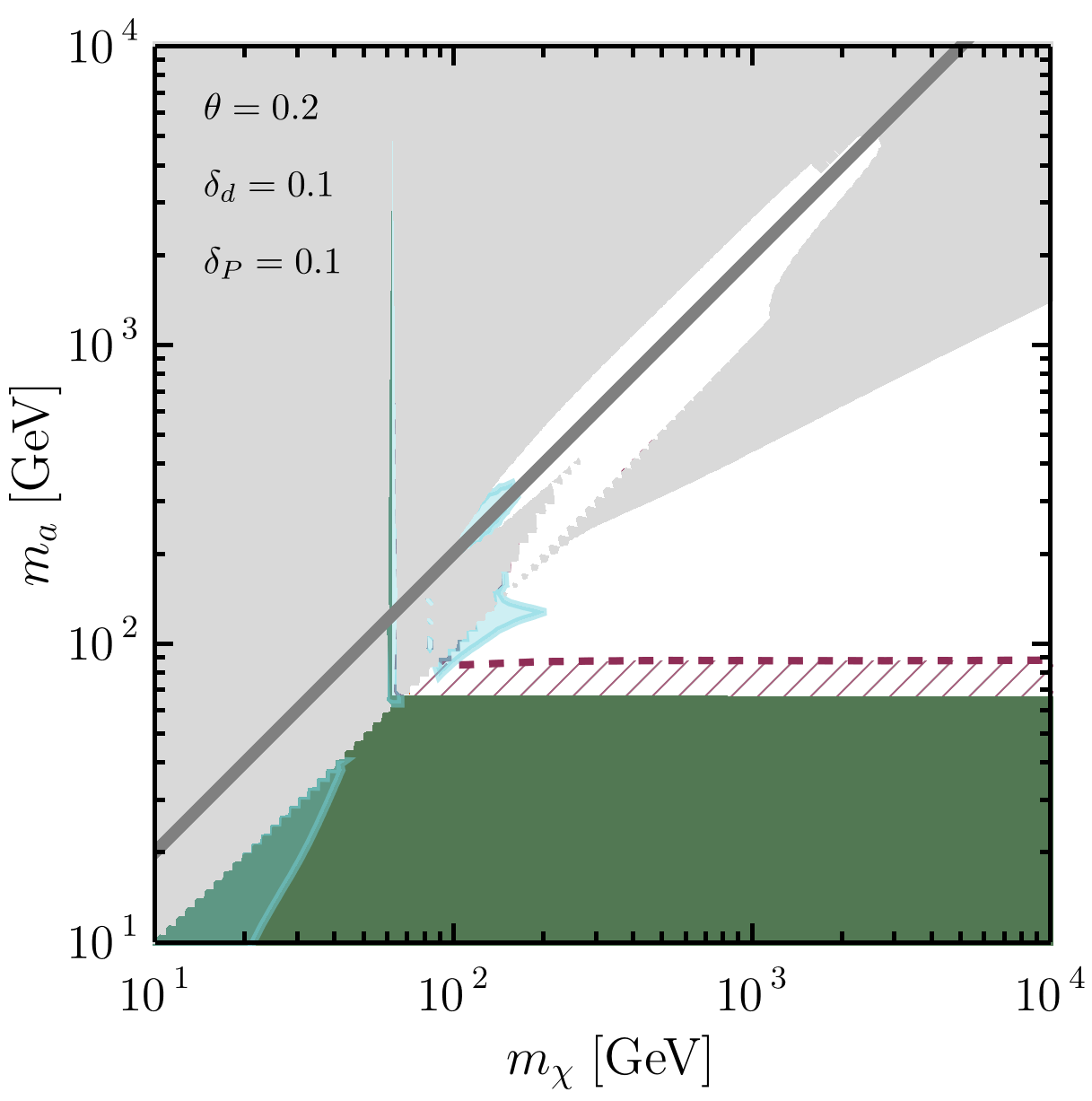}
\end{minipage}\hspace{1.25cm}%
\begin{minipage}[t]{0.39\linewidth}
\includegraphics[width=\linewidth]{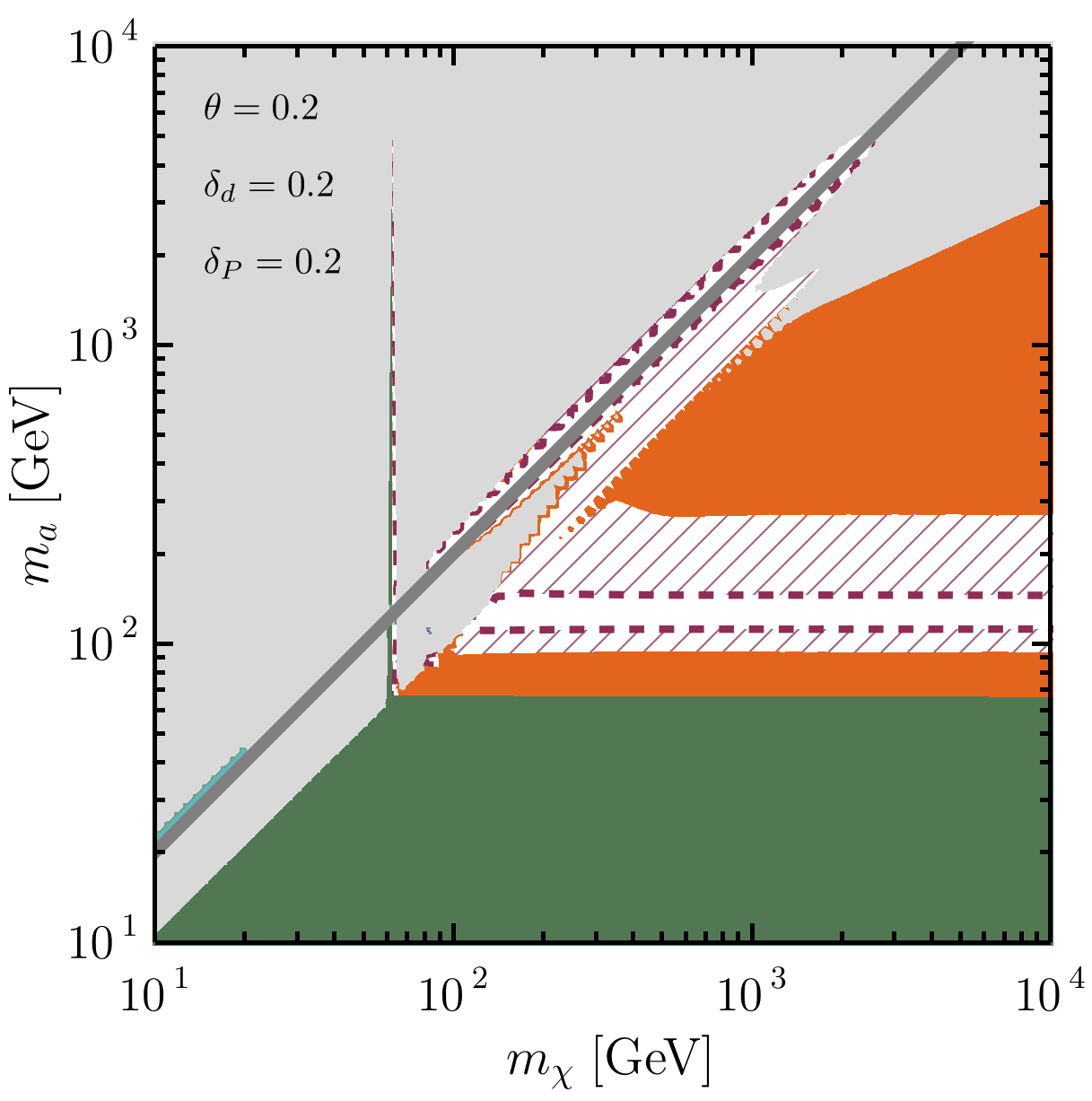}
\end{minipage}\newline\vspace{-0.4cm}%
\caption{\label{fig:Results}  Regions  in the dark sector masses plane where the dark Yukawa coupling required to reproduce the observed DM relic abundance is excluded. Light gray regions are excluded by perturbativity. In green, regions constrained by measurements of the Higgs signal strength~\cite{ATLAS:2022vkf,CMS:2022dwd} and, in orange, by recent results from the direct detection experiment LZ~\cite{LZCollaboration:2024lux}. The red-hatched regions show the sensitivity of the future direct detection experiment DARWIN~\cite{DARWIN:2016hyl}. Excited DM states decaying during and after BBN, $\tau_\ast \geq 0.02$~s~\cite{Boyarsky:2020dzc}, are present in the model for blue shaded regions. The gray bands stand for the resonant region $m_a \in (2\pm10\%)m_\chi$ where kinetic decoupling is expected to alter the relic abundance computation.}
\end{figure*}

In \cref{fig:Results}, we present the combined constraints on the freeze-out parameter region. In each panel, we take different values for the mixing angle $\theta$, the mass ratio $\delta_d$ and the parity breaking parameter $\delta_P$. Then, for each pair of masses of the dark sector particles, the dark Yukawa coupling $y$ is fixed by requiring the correct DM relic abundance to be produced via freeze-out.  Opaque regions are currently excluded by experimental data or perturbativity, while shaded/hatched regions are expected to be excluded by future analyses.  Gray bands represent the resonant region  $m_a \in (2\pm10\%)m_\chi$ where early kinetic decoupling is expected to alter the relic abundance computations~\cite{Brahma:2023psr}. We see that perturbativity imposes strong constraints on most of the parameter space. 

In particular, for visible freeze-out, perturbativity essentially prohibits regions with extremely small $\delta_d \ll 0.1$ forcing the Dirac mass scale $m_d$ to be of similar order to the dark vev $w$ and the elastic interactions to be still relevant for direct detection experiments. This stringent constraint was expected since the dark scalar mass $m_a \sim w \sim \delta_d m_d$ has to be heavier than the DM mass $m_\chi \sim m_d$ in order for DM to be produced via the visible freeze-out mechanism. This implies that, although miDM allows new regions of parameter space in the visible freeze-out regime due to its inelastic nature, the new regions are still very limited by perturbativity and can be nearly fully probed by proposed future direct detection experiments. Only fine-tuned regions where $\delta_P\approx\delta_d$ can completely avoid direct detection constraints. However, fine-tuning works only  for small values of deltas $\delta_i\lesssim 0.2$ since large mass splittings are exponentially suppressed. Last, still in the context of visible freeze-out, we see that small mixing angles $\theta\lesssim 0.1$ are also strongly constrained by perturbativity. Consequently, future searches at colliders may still have the potential to probe important regions of the visible freeze-out parameter space. 

Conversely, in the secluded region, we still have large portions of parameter space allowed by perturbativity since the dark scalar is naturally lighter than the DM particle. Consequently, we have little problem suppressing the direct detection limits by considering ever smaller $\delta_i$. However, even for $\alpha_{\_\_}\approx 0$, constraints from the measurement of the Higgs signal strength are still relevant and can only be avoided by considering extremely small scalar mixing angles $\theta \lesssim 10^{-2}$. Moreover, large parts of the parameter space unconstrained by direct detection or collider searches contain long-lived excited states $\tau_\ast \gtrsim 0.02$~s~\cite{Boyarsky:2020dzc}, due to very small mass splittings. This can affect BBN and CMB data which might effectively rule out those regions as well - we leave the detailed study of miDM cosmology for future works. Considering smaller $\delta_i$ and $\theta$ to relax  Earth-based experimental bounds tend to just increase this region showing the complementarity of cosmological probes in order to fully investigate this type of models.

\section{Conclusion and Outlook}

We have introduced a renormalizable pseudo-Dirac dark matter (DM) model adopting a dark Higgs mechanism with a real pseudoscalar field, which, at the same time, mediates DM-SM interactions via the Higgs portal. The pseudo-Dirac nature is generated via a Majorana-Yukawa term ($\bar{\chi^c}\chi$) by allowing for parity breaking in the dark sector. In the case of small parity breaking,  the diagonal couplings between dark fermions are suppressed, and the model presents an inelastic DM (iDM) behavior. The lack of an extra massive gauge boson, usually present in pseudo-Dirac iDM, establishes the model as one of the most minimalistic iDM models proposed in the literature. 

Now, pseudo-Dirac DM can also be seen as a minimal modification of the standard dark Higgs portal fermionic DM, leading to a qualitatively new cosmological history and novel experimental signatures. In this work, we introduced this new possibility for the first time and examined its properties for large dark DM masses. We focused on the initial conditions independent scenario of DM production via freeze-out with a dark sector in equilibrium with the SM bath. In the visible freeze-out regime, we found that the model is strictly constrained by perturbative unitarity, requiring mediator masses below $m_a\lesssim3 m_\chi$, i.e., some $s$-channel resonant enhancement of the annihilation cross section. In comparison, the standard scenario is essentially ruled out in this regime while the CP-odd portal is not expected to be probed for masses above a few hundreds GeV. 

In contrast, in the secluded regime a larger viable parameter space is present even for large scalar mixing angles $\theta \gtrsim 10^{-2}$ where the standard scenario would be (nearly) completely constrained by direct detection experiments (with the exception of the very fine-tuned region where $m_a \approx m_h$). In this case, the miDM model also allows for lighter DM candidates, since it can avoid indirect detection bounds, in contrast to the CP-odd case. Moreover, it presents regions where the excited DM state can be extremely long-lived promising interesting signatures at both cosmological and collider scales.  

Finally, we point out that the most predictive region of parameter space, where DM is produced via visible freeze-out, will be nearly completely explored (out of the fine-tuned regions where $\delta_P\approx\delta_d \ll 1 $ or $m_a \approx 2m_\chi $) by future direct detection experiments, such as DARWIN. This fact makes the miDM model a particularly interesting target for future direct detection experiments. To conclude, this work sheds light on the important role played by parity violation in the dark sector even for simple models, and invites us to revisit the often assumed parity conservation of our favorite DM models.

\begin{acknowledgments}
The author thanks  (in alphabetical order) Felix Kahlhoefer, Giacomo Landini, Maksym Ovchynnikov and Thomas Schwetz for the fruitful discussions. The author also expresses his gratitude to  Thomas Schwetz for valuable comments on earlier versions of the manuscript and Maksym Ovchynnikov for proof-reading its final version. He also thanks the Doctoral School  ``Karlsruhe School of Elementary and Astroparticle Physics: Science and Technology (KSETA)” for financial support through the GSSP program of the German Academic Exchange Service (DAAD). This work has received support by the European Union’s Framework Programme for Research and Innovation Horizon 2020 under grant H2020-MSCA-ITN-2019/860881-HIDDeN.
\end{acknowledgments}
\appendix

\section{Boltzmann equations}\label{ap:Aboltzmann}

Here, we provide a discussion regarding our DM freeze-out computation. We start from the usual integrated Boltzmann equations for each DM component (assuming all SM particles are in equilibrium with the mediator):  \begin{widetext}
\begin{multline}
    \dot{n}_{\_} +3 H n_{\_} = \langle  \sigma v \rangle_{\_\_} (n_\_^{\text{eq }2}-n_\_^{2}) + \langle  \sigma v \rangle_{\_\ast} (n_\_^{\text{eq}}n_\ast^{\text{eq}}-n_\_ n_\ast ) \\+ 2\langle  \sigma v \rangle_{\chi\chi} \left(n_\ast^{2}-\left(n_\_\dfrac{n_\ast^{\text{eq}}}{n_\_^{\text{eq}}}\right)^2\right) +   (\langle  \sigma v \rangle_{\chi _X}n_X^{\text{eq}}+\Gamma_{\ast}) \left(n_\ast- n_\_ \dfrac{n_\ast^{\text{eq}}}{n_\_^{\text{eq}}}\right) \, , \end{multline} 
    \begin{multline}
    \dot{n}_{\ast} +3 H n_{\ast} = \langle  \sigma v \rangle_{\ast\ast} (n_\ast^{\text{eq }2}-n_\ast^{2}) + \langle  \sigma v \rangle_{\_\ast} (n_\_^{\text{eq}}n_\ast^{\text{eq}}-n_\_ n_\ast) \\ - 2\langle  \sigma v \rangle_{\chi\chi} \left(n_\ast^{2}-\left(n_\_\dfrac{n_\ast^{\text{eq}}}{n_\_^{\text{eq}}}\right)^2\right) -   (\langle  \sigma v \rangle_{\chi _X}n_X^{\text{eq}}+\Gamma_{\ast}) \left(n_\ast- n_\_ \dfrac{n_\ast^{\text{eq}}}{n_\_^{\text{eq}}}\right) \, , \end{multline}
\end{widetext} where $n_{\_(\ast)}$($n_X$) is the number density of $\chi^{(\ast)}$($X$), $H$ is the Hubble rate, and $\langle  \sigma v \rangle_{\_\_}$, $\langle  \sigma v \rangle_{\_\ast} $ , $\langle  \sigma v \rangle_{\chi \chi} $, $\langle  \sigma v \rangle_{\chi _X} $ and $\langle  \sigma v \rangle_{\ast\ast} $ are the thermally averaged cross sections for the processes $\chi \chi \to X Y$, $\chi \chi^\ast \to X Y$ , $\chi^\ast \chi^\ast \to \chi \chi$, $\chi^\ast X \to  \chi X$ and  $\chi^\ast \chi^\ast \to X Y$, respectively, with $X$ and $Y$ representing any SM particle or the pseudoscalar $a$. Assuming that the DM states are in kinetic equilibrium with the SM at a temperature $T$, we can write their ratio as \begin{align} \xi \equiv n_\ast / n_\_ = (1+\Delta_m)^{\frac{3}{2}} e^{-{\Delta_m x}} , \end{align}  with $x = m_\chi/T$. As a result, the integrated Boltzmann equations can be simplified into \begin{equation}\label{eq:MainBoltzmainEquation}
    \dot{n} +3 H n = \langle  \sigma v \rangle (n^{\text{eq }2}-n^{2})  
\end{equation} where $n=n_\_+n_\ast$ and \begin{equation}
    \langle  \sigma v \rangle = \dfrac{ \langle  \sigma v \rangle_{\_\_}}{(1+\xi)^2}+2\xi\dfrac{  \langle  \sigma v \rangle_{\_\ast}}{(1+\xi)^2}+\xi^2\dfrac{  \langle  \sigma v \rangle_{\ast\ast}}{(1+\xi)^2} \, .
\end{equation} This is equivalent to solving the Boltzmann equation for a single DM species. 

Since $\Delta_m \ll 1$, we approximate $m_{\chi^\ast} \approx m_{\chi}$  in all thermally averaged cross sections. Moreover, we velocity-expand our cross sections up to $d$-wave ($v_\chi^4$), using the analytical formulas given in ref.~\cite{Wells:1994qy}, in order to speed up our computations. Finally, we numerically solve \cref{eq:MainBoltzmainEquation} using \texttt{Mathematica} as well as using the usual freeze-out approximation. Both solutions present good agreement with results found in the literature for classical WIMP models as well as the standard dark photon-mediated iDM scenario.

\bibliography{main}% Produces the bibliography via BibTeX.

\end{document}